\documentclass[pdflatex,sn-basic,Numbered]{sn-jnl}

\usepackage{graphicx}%
\usepackage{multirow}%
\usepackage{amsmath,amssymb,amsfonts}%
\usepackage{bm}%
\usepackage[title]{appendix}%
\usepackage{xcolor}%
\usepackage{textcomp}%
\usepackage{booktabs}%
\usepackage{array}%
\usepackage{url}%
\usepackage{csquotes}%
\usepackage{rotating}%

\newcommand{\ctext}[1]{\raise0.2ex\hbox{\textcircled{\scriptsize{#1}}}}
\newcommand{\qualparagraph}[1]{%
  \par\addvspace{1.5ex}\noindent\textbf{#1}\par
  \noindent
}

\begin{document}

\title[Towards Restorative Urban Spaces]{Towards Restorative Urban Spaces:\\An Exploratory Field Study of Nature Videos on Public Multi-Display Signage}

\author*[1]{\fnm{Yuri} \sur{Mikawa}}\email{yuri\_mikawa@ipc.i.u-tokyo.ac.jp}

\author[2]{\fnm{Takumi} \sur{Yokosaka}}

\author[2]{\fnm{Yuki} \sur{Kubota}}

\author[2]{\fnm{Kazushi} \sur{Maruya}}

\affil*[1]{\orgdiv{Graduate School of Information Science and Technology}, \orgname{The University of Tokyo}, \orgaddress{\city{Tokyo}, \postcode{113-8656}, \country{Japan}}}

\affil[2]{\orgdiv{Communication Science Laboratories}, \orgname{NTT Inc.}, \orgaddress{\city{Kanagawa}, \postcode{243-0198}, \country{Japan}}}

\abstract{In recent years, multi-display signage (MDS) has become increasingly common in urban stations, plazas, and other public spaces, where it is primarily used to present advertisements and information. Beyond these conventional applications, MDS offers opportunities to shape people's experiences of public environments through spatially distributed and dynamic visual content. Motivated by the potential of mediated nature to support restorative experiences in urban environments, we conducted a nine-day exploratory field study in a free-admission area of a science museum, presenting artificial intelligence (AI)-generated nature videos across four displays and collecting 1,263 questionnaire responses. We combined three complementary analyses. First, structural equation modeling identified strong associations among perceived consistency across displays, presence, and self-reported restorative responses, consistent with our theoretically motivated model. Second, exploratory comparisons across video-presentation configurations revealed modest differences in consistency and restoration ratings, although these AI-generated videos did not significantly outperform the advertisement reference. Third, qualitative analysis of open-ended responses identified fascination, immersion, and restorative feelings, alongside concerns about naturalness, visual quality, and motion-related discomfort. Together, these findings provide empirical insights into MDS-mediated nature experiences, with implications for nature-video design and questionnaire refinement. They also inform a tentative conceptual extension of the model incorporating additional perceptual and evaluative factors.}


\maketitle

\section{Introduction}

\begin{figure*}
    \centering
    \includegraphics[width=\linewidth]{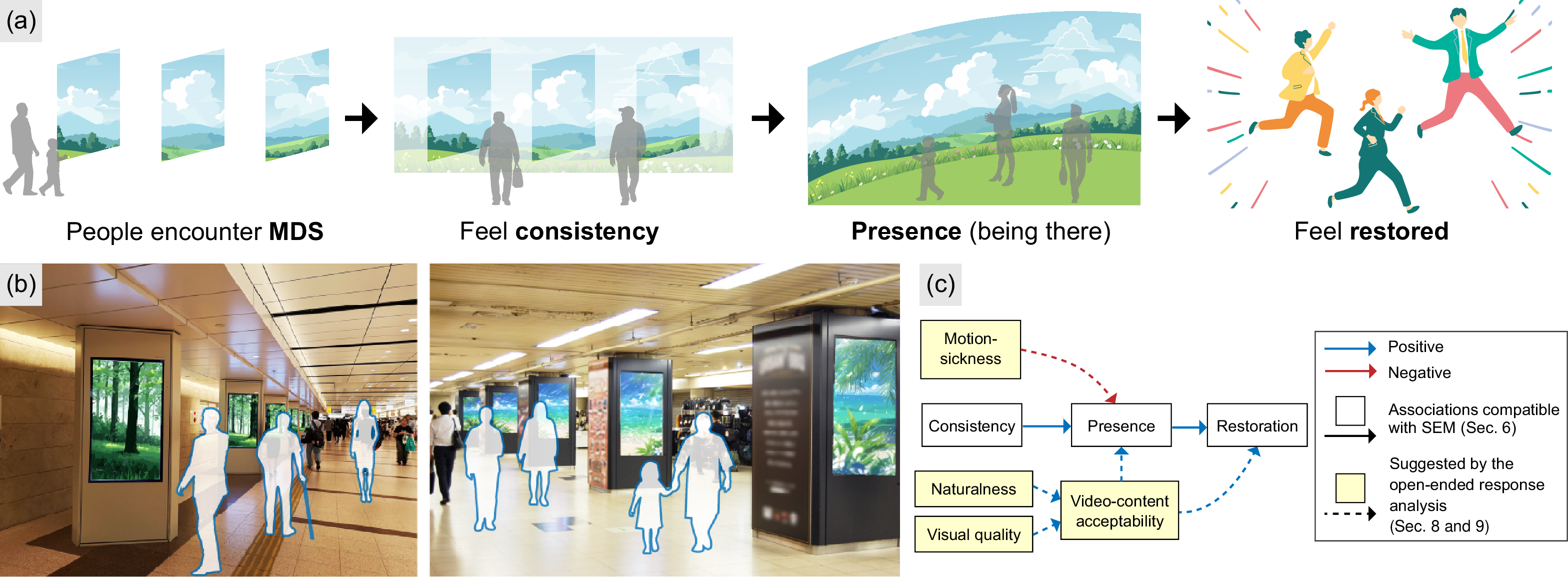}
    \caption{(a) Hypothesized three-factor model of MDS-mediated restorative responses, in which theoretically motivated directional paths link consistency, presence, and restoration. (b) Conceptual image of nature imagery presented across an MDS in a station, created by overlaying copyright-free nature videos onto a copyright-free photograph of an actual urban environment. (c) Conceptual extension of the model, derived post hoc from the integrated observations across the three analyses, particularly the open-ended responses}
    \label{fig:teaser}
\end{figure*}

In contemporary urban environments, it is increasingly common to find multi-display signage (MDS) installed throughout public spaces. At present, these displays mainly function as repetitive advertising media or information boards. However, such spatially distributed signage systems can also be regarded as a form of spatial display that is uniquely capable of presenting dynamic visual content across public space. For example, they may enable visual expressions that evoke the impression of looking into another world beyond the signage, thereby potentially fostering feelings of immersion, escape, and extraordinariness.

People's engagement with urban signage is often non-interactive and brief, particularly in the case of MDS. This type of experience reflects how people frequently encounter signage in crowded urban settings, such as by casually noticing content in their peripheral vision while passing by or by momentarily becoming aware of it and forming a brief impression. However, previous studies on public signage have mainly focused on single, continuous display surfaces and on interactive scenarios in which people actively engage with signage through embedded user interfaces or explicit interaction techniques \cite{muller2010requirements,michelis2011audience,parra2014understanding,veenstra2015should,wouters2016uncovering}. In contrast, relatively little research attention has been paid to non-interactive experiences in which people simply view signage content while passing by, let alone MDS \cite{memarovic2012using}.

A promising form of such non-interactive experience with MDS is the presentation of nature videos. In contemporary crowded urban environments, opportunities for psychological restoration are becoming increasingly important. Urban commuting and public transportation can be sources of psychological and physiological stress, with prior studies linking commuting duration, unpredictability, and crowding-related personal-space invasion to perceived stress and stress responses \cite{evans2002morning,evans2006rail,evans2007crowding,norgate2020public}. In contrast, natural environments and mediated nature, including photographs, videos, and virtual reality (VR), have been shown to support psychological restoration and stress recovery \cite{ulrich1991stress,berto2005exposure,de2006s,valtchanov2010restorative,li2023effects}. These findings motivate the use of MDS as an urban-scale medium for introducing restorative natural content into crowded spaces where direct access to real nature is limited, as shown in Fig. \ref{fig:teaser}(b).

Motivated by this possibility, the present study explores whether, how, and to what extent passers-by report restorative responses to nature videos presented on MDS in public environments. We conducted a field study using four vertically oriented displays installed in a free-admission area of a museum, providing a setting that shares characteristics with MDS installations in stations, plazas, and other public spaces. Staff members briefly directed visitors' attention to the displays and invited them to complete a questionnaire after passing through the area. Viewing circumstances varied across participants, including their social context, walking behavior, viewing distance, and age. Under these diverse conditions, we investigate how people experience MDS-mediated nature while walking through a public space.

To achieve the objective of this study, we proceed in three steps. First, we propose a simple, theoretically motivated model of MDS-mediated restorative responses (Fig. \ref{fig:teaser}(a)), construct a questionnaire based on this model, and use confirmatory factor analysis (CFA) and structural equation modeling (SEM) to examine whether the hypothesized structure is compatible with responses collected in the field (Sec. \ref{sec:validation-model}). Because the data are cross-sectional self-reports, this analysis examines associations among the three constructs rather than causal or mediating processes.
Second, we compare responses across complete MDS presentation configurations that differed in viewpoint motion, segment arrangement, and display spacing, together with the museum's existing advertisement video as a reference condition (Sec. \ref{sec:stat-analysis}). Because each configuration was presented on a single day for each spacing level, these comparisons are exploratory and cannot isolate the effects of individual design features from day-level factors.
Third, we analyze optional open-ended responses to explore participants' impressions beyond the rating-scale items (Sec. \ref{sec:open-text}). We then use these results to discuss possible design considerations, to propose a hypothesis-generating extension of the model (Fig. \ref{fig:teaser}(c)), and to derive implications for nature-video and questionnaire design (Sec. \ref{sec:consideration}).

This study takes an exploratory but systematic approach to understanding restorative responses to nature videos presented on MDS in a real-world public setting. Using 1,263 questionnaire responses collected over nine days at a science museum, we examine associations among visitors' ratings of consistency, presence, and restoration, differences in ratings across presentation configurations, and the experiences described in their comments. Together, these analyses provide insights into how people experience MDS-mediated nature and identify important considerations for designing nature videos and evaluating restorative responses in public spaces.

\section{Related work}\label{sec:related-work}

\subsection{Restoration theory}\label{sec:restoration-theory}
Two major frameworks for explaining restorative experiences are Attention Restoration Theory (ART) \cite{kaplan1989experience,kaplan1995restorative} and Stress Recovery Theory (SRT) \cite{ulrich1991stress}. ART describes restoration as the replenishment of depleted directed attention through four key characteristics---\textit{being away}, \textit{extent}, \textit{soft fascination}, and \textit{compatibility}---whereas SRT provides a more psychophysiological account of affective and physiological stress recovery from natural environments \cite{ohly2016attention}. ART is especially relevant to the present study because the MDS first functions to attract viewers' attention through large-scale visual presentation. This relevance is also reflected in the open-ended response analysis (Sec. \ref{sec:open-text}).

Although these relationships have not been directly examined in the context of MDS, prior research provides partial support for the proposed links among consistency, presence, and restoration. Perceptual coherence and consistency in mediated environments can enhance presence by making the environment feel unified and plausible \cite{slater2009place,skarbez2017survey,skarbez2017psychophysical,skarbez2021immersion,lim2023effects}, and presence has been associated with stronger restorative responses in virtual or simulated nature \cite{de2006s,valtchanov2010restorative,mattila2020restoration,spano2023virtual}. In addition, the possible relevance of consistency to restoration is consistent with ART, in which coherence and compatibility are regarded as key characteristics of restorative environments \cite{kaplan1989experience,kaplan1995restorative,ohly2016attention}. Thus, prior studies provide theoretical support for individual links, such as consistency--presence and presence--restoration relationships. In contrast, the present study proposes an integrated psychological model for MDS-mediated restorative experiences that incorporates consistency, presence, and restoration within a single framework.

\subsection{Restorative experience using natural environments and digitally mediated nature} 

Research on restorative experiences has addressed both direct exposure to physical natural environments and indirect exposure through visual media. Natural environments are generally perceived as more restorative than urban environments, and systematic reviews have reported beneficial effects of exposure to nature on concentration, mood, and mental recovery \cite{menardo2021restorativeness}. However, this contrast should not be interpreted as indicating that urban environments are intrinsically non-restorative. Previous studies have shown that well-designed or aesthetically valued urban landscapes can also afford restorative benefits comparable to those of natural environments \cite{karmanov2008assessing,sanjuan2017restoration}. These findings suggest that environmental restoration is determined not solely by the categorical distinction between \enquote{natural} and \enquote{urban} settings, but also by their visual, spatial, and experiential qualities.

Restorative effects have also been reported for mediated nature, including photographs, projected imagery, broadcast video, television, and virtual reality. Early work showed that even visual representations of nature can support recovery from stress and attentional fatigue, suggesting that restorative experiences are not limited to direct physical contact with natural environments \cite{ulrich1991stress,berto2005exposure}. De Kort et al. further demonstrated that stress recovery can occur in mediated natural environments and examined how the level of immersion influences this process \cite{de2006s}. Subsequent studies have extended this line of work to virtual and simulated nature, showing that VR forests and other computer-generated natural environments can improve perceived restoration, vitality, mood, stress reduction, and emotional arousal \cite{valtchanov2010restorative,mattila2020restoration,li2022virtual}. Recent reviews likewise indicate that virtual nature commonly produces benefits in restorativeness, although the magnitude of these effects depends on factors such as presentation quality, immersion, interaction, and study design \cite{spano2023virtual,li2023effects}.

Together, these findings suggest that nature's restorative potential can be elicited not only through real environments but also through digital media. Building on this perspective, the present study examines restorative experiences in which digital nature presented through MDS is intermingled with the surrounding urban environment. To the best of our knowledge, this form of MDS-mediated nature experience has not been directly investigated in prior research. It further explores whether restorative responses are reported after brief, non-interactive viewing of such digital nature while people walk through a space, rather than after deliberate, sustained viewing in controlled or immersive settings.

\subsection{Public displays in urban space} 

Research on public displays has increasingly emphasized the importance of in-the-wild field studies in real urban settings, where pedestrian flow, incidental attention, and situated social behavior cannot be fully reproduced in laboratory environments \cite{alt2012evaluate}. Prior work has therefore examined how public displays attract passers-by, overcome display blindness, encourage stopping and participation, and shape interaction around the display. For example, previous studies have examined how people's expectations about displayed content influence their attention to digital signage \cite{muller2009display}, how passers-by move through stages of attention and interaction around gesture-based displays \cite{michelis2011audience}, and how visual feedback or interactive cues can make a public display more noticeable \cite{brignull2003enticing,muller2012looking}. Large-scale deployments in city centers have also demonstrated the value of public displays as long-term urban computing infrastructures for studying real users and everyday situated behavior \cite{ojala2012multipurpose}.

A smaller but related body of work has explored how public or interactive displays can influence people's affective and social experiences. Studies of affective and interactive installations such as Mood Swings and {\itshape mood.cloud} have examined how public media can visualize, express, or share emotional states \cite{bialoskorski2009mood,scolere2016building}, while systems such as Blobulous have investigated whether publicly displayed interactive agents can support social connectedness in public spaces \cite{funk2013connected}. Interactive technologies have also been studied as a means of reducing anxiety in specific public settings, such as children's hospitals \cite{lim2019designing}.
However, these studies primarily relied on active user engagement with public displays to support affective and social experiences.
In contrast, the present study investigates whether non-interactive nature video content presented across MDS is associated with self-reported restorative responses among passers-by in a public field setting. 

\section{MDS restoration model}\label{sec:MDS-restorative-model}

This section explains a proposed MDS restoration model, as conceptualized in Fig. \ref{fig:teaser}(a). The model consists of three factors that describe passers-by's experiences while walking through a space in which nature imagery is presented on MDS. They are defined as follows.
\begin{itemize}
    \item \textbf{Consistency} refers to the perceived continuity, or cognitively inferred continuity, of imagery across physically separated signage displays, such that the visual content appears to extend beyond the boundaries of individual screens. This impression is supported by homogeneous imagery and temporally synchronized motion across the MDS.
    \item \textbf{Presence} refers to the subjective sense of being situated within, or moving through, the image space presented by the MDS imagery, rather than merely observing it from outside. It encompasses the feeling of being inside the depicted space, reduced awareness of the surrounding real space while walking, and the lingering impression that one had been present in the depicted space after viewing the imagery.
    \item \textbf{Restoration} refers to the self-reported restorative outcomes that viewers attribute to the natural scenery presented through the MDS imagery. It encompasses feelings of calmness, refreshment, renewed energy and motivation, improved attentional readiness toward the surroundings, reduced concern or worry, and mental refreshment after experiencing the imagery. 
\end{itemize}

We developed a sequential restoration model by placing the three factors in theoretically motivated roles: consistency as a perceptual appraisal, presence as an experiential state, and restoration as an affective and cognitive outcome. On this basis, we hypothesize directional paths from consistency to presence and from presence to restoration. 
This interpretation is partially supported by prior studies, although they differ from the present study in how consistency is defined and in the medium through which presence is experienced (see Sec. \ref{sec:restoration-theory}). However, prior research has not examined how the fragmented and distributed presentation of natural imagery on MDS shapes restorative experience through perceptual consistency and presence. The present model is therefore introduced as an MDS-specific framework for examining how consistency, presence, and restoration may be structurally related in this particular form of mediated nature.

\subsection{Questionnaire design}
To examine the participants' perceived consistency, presence, and restoration based on our MDS restoration model, an 11-item questionnaire was designed. To reduce visitor burden and discomfort in the museum, it was necessary to limit the number of questionnaire items. Also, because of the exploratory nature of this study, we constructed the questionnaire by combining established scales with original items developed for consistency.

Because the study placed particular emphasis on the assessment of restorative responses, we used the six-item Restorative Outcome Scale (ROS) \cite{korpela2008determinants,korpela2010favorite}, with the item wording adapted to refer to the experience of the presented images (Table S1 in Online Resource 1). For presence, we selected three items from the six-item Slater--Usoh--Steed (SUS) questionnaire that specifically capture the sense of being inside the depicted image space \cite{usoh2000using}. For consistency, we developed two original items. The questionnaire was composed of 11 items in total: 2 items for consistency, 3 items for presence, and 6 items for restoration. This was followed by an open-ended question in which participants were asked to freely describe their impressions of the nature presented through the MDS, as well as a demographic questionnaire on age group and gender. The questionnaire used in this study is shown in Online Resource 1 (Section S2, Table S1).
The adjective-based or short descriptive labels used to denote each questionnaire item are shown below.

\begin{itemize}
    \item Consistency (Q1, Q2): continuous, expanding
    \item Presence (Q3-Q5): being-there, easy-to-walk, existing
    \item Restoration (Q6-Q11): calm, fine, restored, focused, stress-free, clear
\end{itemize}

\section{Nature video design on MDS}

This section explains how to select and organize the natural landscape imagery for MDS. This study presents a natural landscape spanning multiple displays to promote consistency across displays, rather than conventional repetitive content. A wide range of visual materials can be used, including real-world footage captured with a conventional aspect ratio, such as 16:9 widescreen video; cinemascopic footage of natural landscapes; and computer-generated (CG) open-world imagery that supports 360$^\circ$ viewing, created either by humans or by artificial intelligence (AI).

Among these options, we adopted AI-generated videos created from wide natural landscape images.
Because the MDS presents spatially separated images, it is difficult for viewers to perceive the continuity of the overall scene. We therefore selected natural images with carefully composed layouts, in which key elements supporting scene understanding were effectively arranged within the frame.
In addition, generative AI makes it possible to create videos with controlled viewpoint motion from such natural images, allowing us to vary viewpoint motion while keeping the depicted scenes constant.

However, AI-generated videos have several drawbacks, such as an artificial appearance and low image resolution, even with current cutting-edge AI technologies, which may make the scenes appear unnatural and diminish viewers' sense of presence and restoration. We address these concerns, together with the positive aspects, in the following experiment, and further discuss them by comparing the quantitative evaluation results with the open-ended responses (see Secs. \ref{sec:open-text} and \ref{sec:consideration-compare-open-ended}).

\section{Experimental design}\label{sec:experimental-design}

\begin{figure}
    \centering
    \includegraphics[width=0.5\linewidth]{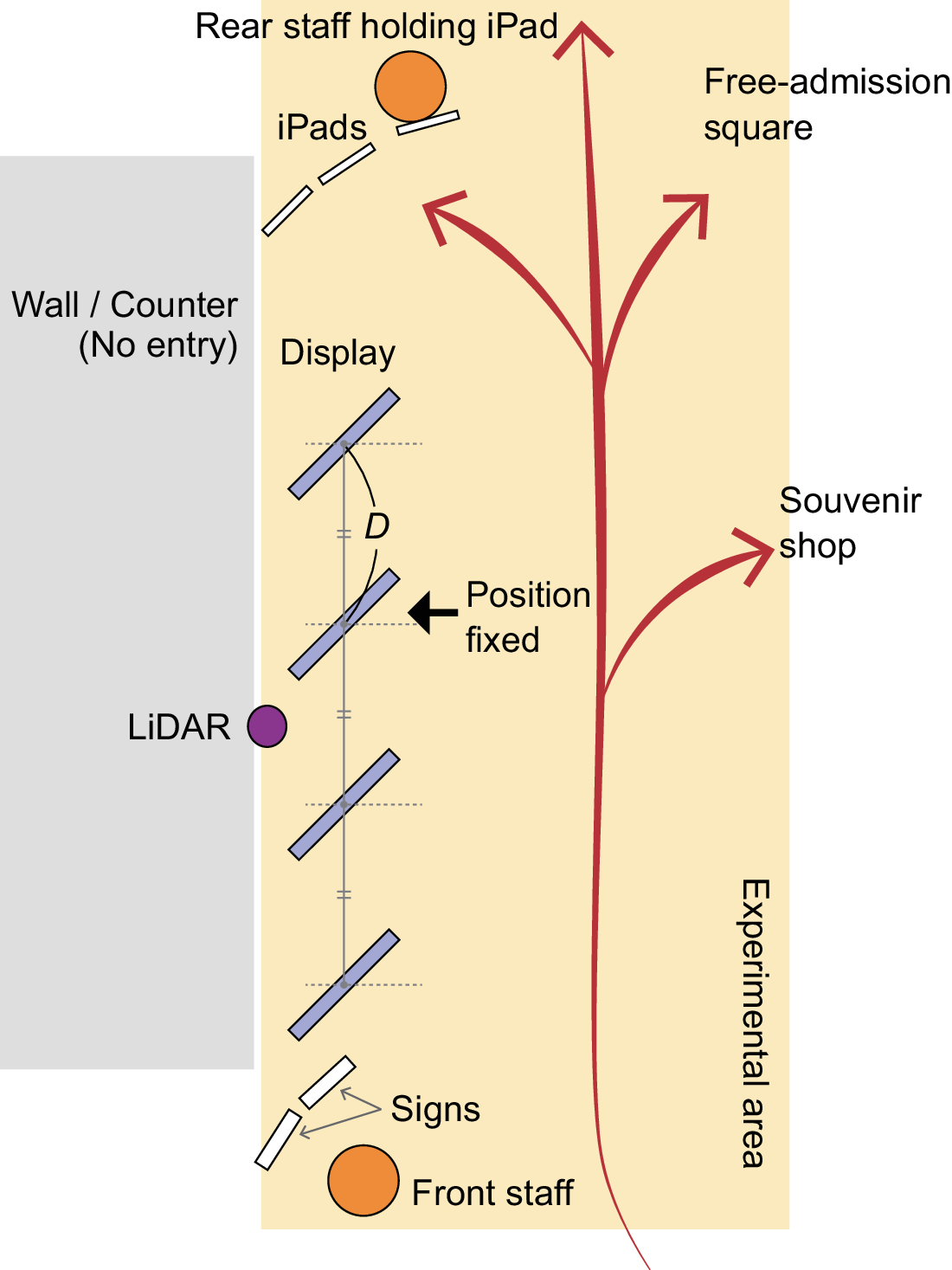}
    \caption{Layout of the experimental space in the National Museum of Emerging Science and Innovation. The red arrows indicate the main pedestrian flow of visitors. The two signboards at the entrance explain the purpose of the experiment and indicate that photography is prohibited, respectively}
    \label{fig:layout}
\end{figure}
\subsection{Experimental setup}\label{sec:experimental-setup}

\begin{figure*}
    \centering
    \includegraphics[width=\linewidth]{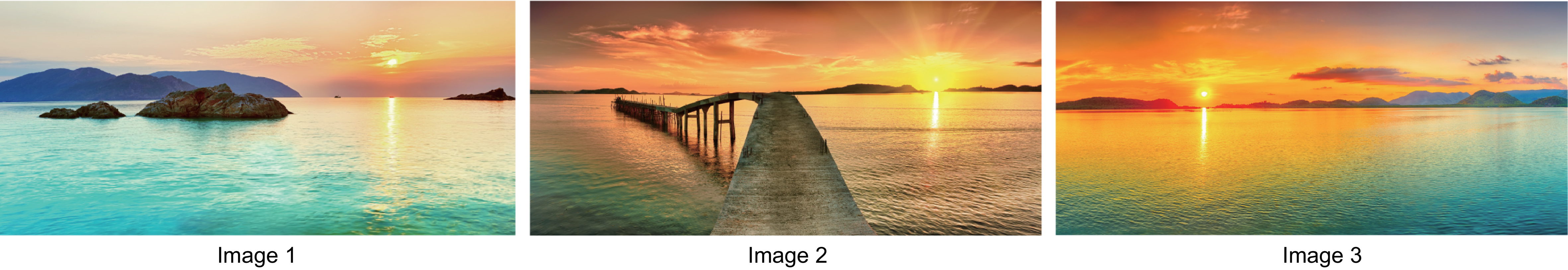}
    \caption{Three source landscape images obtained from Adobe Stock and cropped to a common aspect ratio of 9:4 for video generation}
    \label{fig:nature-images}
\end{figure*}
The experimental layout and the main pedestrian flow are shown in Fig. \ref{fig:layout}. The experiment was conducted near the entrance of the National Museum of Emerging Science and Innovation, Japan. After entering the museum, visitors encountered the experimental area directly ahead and were invited to experience the MDS-mediated nature imagery. At the midpoint of the experimental area, or after passing through it, the pedestrian flow diverged into multiple routes, including routes toward the questionnaire tablets, the souvenir shop, and the free-admission public area. In Fig. \ref{fig:layout}, these forward pedestrian routes are indicated by red arrows. The area tended to be more crowded on weekends and holidays than on weekdays. The experimental area was located in an atrium-like space, where ample natural light entered from outside. 

In the experimental area, four 55-inch vertically oriented displays (JAPANNEXT JN-IPS5503TUHDR, 9:16, 1080$\times$1920 px, IPS panel) were installed as shown in Fig. \ref{fig:layout}. The height from the floor to the center of each display was approximately 1.4 m. Each display was rotated by 45 degrees around the vertical axis so that it faced the pedestrian flow. The displays were arranged along the depth direction with a center-to-center spacing of \textit{D}. This spacing between displays was adjusted while keeping the position of the second display from the back fixed. All displays were connected to a single desktop PC (CPU: Intel Core(TM) Ultra 9 285K, GPU: two NVIDIA RTX A1000 8GB), and the videos were rendered using OpenGL 4.6.0 with a target frame rate of 24 fps.

Two staff members were stationed at the front and rear of the experimental area; during some periods, only the rear staff member was present. In a preliminary experiment, many visitors passed through the area without noticing that an experimental demonstration was being conducted. Therefore, in the present experiment, the front staff member verbally drew visitors' attention to the MDS, while the rear staff member encouraged them to complete the questionnaire on tablets after they had passed through the area. Three iPads were prepared for questionnaire responses: two (Apple MD4A4J/A, 11") were mounted on stands in the area beyond the passage route, and the rear staff member carried the third (Apple MPQ03J/A, 10.9"). The questionnaire was created using Typeform \cite{typeform}, which provides an intuitive interface that allows participants to answer questions smoothly and easily across a wide range of ages. It was presented solely on the iPads, and no public link was distributed online, thereby restricting it to only on-site responses. 

To confirm the presence of pedestrian flow, a LiDAR sensor (Hesai Technology JT128) was installed on the ceiling between the second and third displays. However, the resulting point clouds were sparse, and we judged that they lacked sufficient detail to estimate features such as visitors' head orientation. These data were therefore not analyzed. As shown at the bottom of Fig. \ref{fig:layout}, signs were placed at the front of the experimental area providing information regarding privacy protection and consent regarding the collection of point-cloud data using LiDAR, as well as a notice prohibiting photography.

\subsection{Stimuli}\label{sec:stimuli}
We searched Adobe Stock for sunrise and sunset landscape images and obtained three images by the same creator, Olga Khoroshunova (Images 1--3). Their original resolutions were 6000$\times$2357, 7136$\times$3277, and 14005$\times$3534 px, respectively. Each image was cropped to an overall aspect ratio of 9:4 so that its four equally sized segments would each have the 9:16 aspect ratio of a portrait display (9$\times$4:16 = 36:16 = 9:4). The resulting images are shown in Fig. \ref{fig:nature-images}.

Eight-second, 24-fps videos with motion were generated from each image using Adobe Firefly (model: Google Veo 3.1). For each image, two video types were generated: one in which the viewpoint remained fixed (Fixed) and one in which the viewpoint continuously moved forward into the scene (Moving). For each movement type, the three videos generated from the three images were then temporally concatenated in the order of Images 1--3, corresponding to the left-to-right order shown in Fig. \ref{fig:nature-images}, and played in a loop. Optical-flow analysis (Online Resource 1, Section S1, Fig. S1) indicated that the moving-viewpoint videos contained substantially stronger visual motion than the fixed-viewpoint videos, with mean angular velocities of 3.42$\pm$1.28$^\circ$/s and 0.77$\pm$0.35$^\circ$/s, respectively, assuming a viewing distance of 2.0 m and a frame rate of 24 fps. The visual motion in the moving-viewpoint videos overlapped with the lower end of the range of angular velocities used in previous vection studies, including 2--64$^\circ$/s for roll circular vection and 5--60$^\circ$/s for visually induced yaw vection \cite{wang2021visually,nesti2015self}.

The resolution of the generated videos from Adobe Firefly was 1920$\times$1080 px. Each video was divided horizontally into four equal segments and presented across four displays, resulting in an effective resolution on each display of 480 px in width and 1080 px in height.
The luminance at the center of the second panel from the right among the four divided video panels was measured using a luminance meter (KONICA MINOLTA CS-150) with a measurement angle of 1\textdegree{}. Across 100 measurements taken over 344.3 s (approximately 14.3 video loops), the luminance was 150.46 $\pm$ 45.97 cd/m$^2$ (mean $\pm$ standard deviation).

\begin{figure*}
    \centering
    \includegraphics[width=\linewidth]{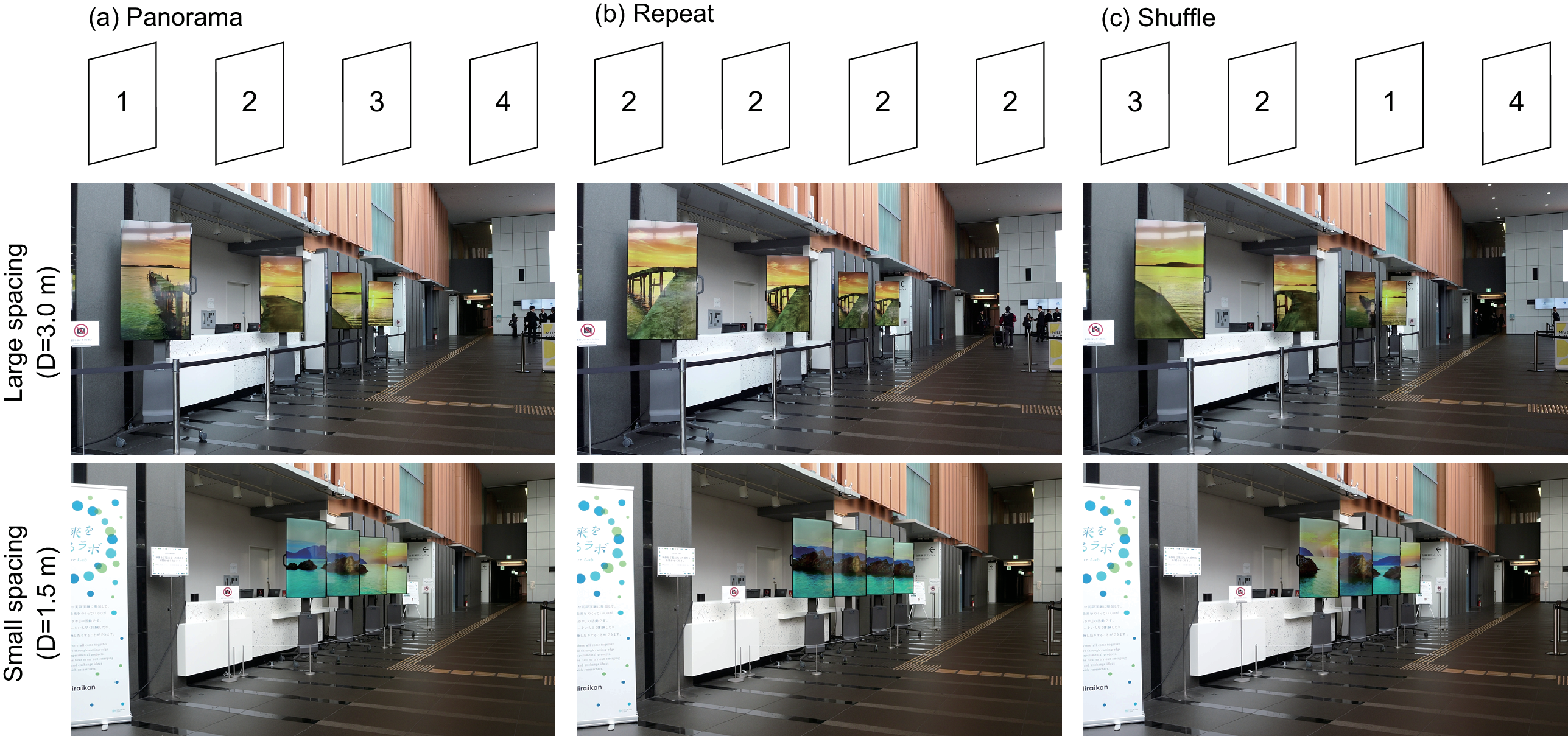}
    \caption{Three types of image ordering in MDS used for the experiment, and the appearances presented in the actual MDS with large ($D=3.0$ m) and small ($D=1.5$ m) spacing. The position of the second display from the back was always fixed. The displays show nature-video stimuli generated from source images using Adobe Firefly (Google Veo 3.1; Sec. \ref{sec:stimuli})}
    \label{fig:img-order}
\end{figure*}

To examine whether consistent visual connections between displays contribute to visitors' experience and perceived restoration, three image placements across the four displays were used, as shown in Fig. \ref{fig:img-order}(a--c): panorama, repeat, and shuffle. In the panorama condition, adjacent video segments were assigned to neighboring displays so that the spatial continuity of the original panorama was preserved across the MDS. In the repeat condition, the same video segment of the second display in the panorama condition was displayed across all displays. In the shuffle condition, the mapping of video segments to displays was randomized relative to their original panoramic sequence, thereby disrupting spatial continuity across the displays. The same randomized mapping was used across the spacing conditions.

The spacing between the signage displays was also manipulated, with a large-spacing condition ($D=3.0$ m) and a small-spacing condition ($D=1.5$ m), as shown in Fig. \ref{fig:img-order}. This manipulation was used to examine whether greater separation between video segments, as commonly observed in urban MDS installations, would reduce perceived consistency across the displays and influence viewers' experiences and restorative responses.
In the large-spacing condition, the gap between adjacent displays was approximately equivalent to the width of a signage display, whereas in the small-spacing condition, there was almost no physical gap between displays. However, because the displays were angled relative to one another, the visual continuity of the imagery between adjacent screens was not fully preserved.

The videos themselves and the appearance of the MDS displaying the nature images are also shown in the supplementary video (Online Resource 2).

\subsection{Conditions}\label{sec:condition}

\begin{figure}[h]
    \centering
    \includegraphics[width=0.5\linewidth]{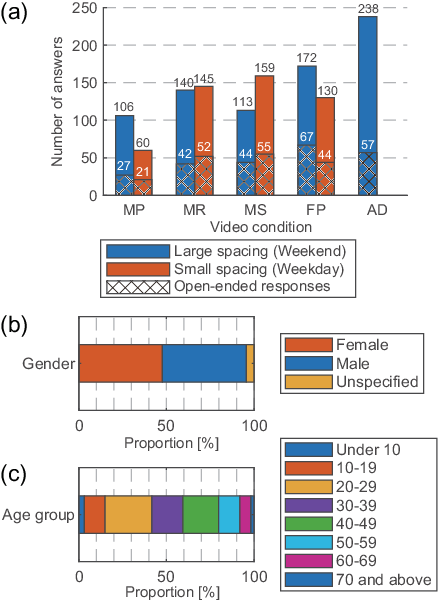}
    \caption{(a) Number of questionnaire responses and open-ended responses for each condition (each day). The numerical values are shown above each bar. Proportions of respondents by (b) gender and (c) age group, aggregated across all responses}
    \label{fig:answer-num}
\end{figure}

The experiment employed a between-subjects field design in which each visitor experienced the configuration presented on the day of their visit.
Center-to-center display spacing had two levels: large ($D=3.0$ m) and small ($D=1.5$ m).
As nature-video configurations, three image-order configurations, moving-panorama (MP), moving-repeat (MR), and moving-shuffle (MS), were prepared for the forward-moving viewpoint. In addition, fixed-panorama (FP) was prepared as a fixed-viewpoint configuration. Among these, MP was designed as the main configuration combining viewpoint-motion cues with spatial continuity across the displays. MR and MS used the same type of moving nature videos but reduced the spatial-continuity cues by repeating or shuffling the video segments, respectively. FP preserved the panoramic spatial arrangement but removed the viewpoint-motion cue. The museum's existing advertisement video (AD) was also included as a reference condition representing conventional signage content. Thus, five video configurations were prepared in total.


The experiment lasted for nine days in total, including five weekend days and four weekdays. Because the number of museum visitors was expected to differ substantially between weekdays and weekends, the experimental schedule was arranged to roughly balance the number of respondents across video configurations within the available experimental period. Specifically, the large-spacing condition was tested on the five weekend days with the MP, MR, MS, FP, and AD configurations, whereas the small-spacing condition was tested on the four weekdays with the MP, MR, MS, and FP configurations. Each configuration--spacing combination was therefore presented on a single day.
The study focused primarily on spatial arrangement and viewpoint motion in the nature videos, with AD included as a reference only under large spacing.


Display spacing coincided with weekday/weekend scheduling, and weather varied across days; these contextual differences are discussed in Sec. \ref{sec:limitation-allocation}. Weather was not included as an analysis factor.

\subsection{Participants}

This experiment targeted all visitors who came to the museum and passed through the experimental area, and those who responded to the final questionnaire were treated as the participants in the experiment. An information notice describing the experimental objective, the measurements to be conducted, the safety of laser exposure from the LiDAR, and the fact that passage through the area would be regarded as consent was posted in front of the area (see the bottom of Fig. \ref{fig:layout}). Visitors were assumed to have consented by passing through the area, and those who did not consent were allowed to opt out by informing the on-site staff (in practice, no one requested to opt out). The questionnaire also stated at the beginning that consent to provide personal information, namely age group and gender, would be deemed given by completing the questionnaire.
The experimental protocol was approved by the research ethics committee of NTT, Inc. and conducted in accordance with the ethical standards of the Declaration of Helsinki.

The aggregated demographic data from the questionnaire responses are shown in Fig. \ref{fig:answer-num}. In total, 1,263 participants completed all required questionnaire items and were included in the analysis. No responses contained errors, and therefore no data were excluded.
Of the respondents, 47.74\% reported their gender as female, 47.90\% as male, and 4.35\% did not specify their gender. The age-group distribution was as follows: under 10 (3.25\%), 10--19 (11.80\%), 20--29 (26.60\%), 30--39 (17.74\%), 40--49 (20.35\%), 50--59 (12.19\%), 60--69 (6.25\%), and 70 and over (1.82\%). Although the sample was somewhat skewed toward younger adults, particularly those in their 20s, the age distribution remained relatively broad.

Although participants' nationalities were not collected, the questionnaire was translated into the local language of the site. Thus, while their nationalities cannot be identified, the participants were likely speakers of, or at least able to read, the corresponding language.

\subsection{Procedure}\label{sec:procedure}
A staff member positioned in front of the experimental area verbally informed visitors about the experiment using the multi-display signage. Visitors then passed through the experimental area; some visitors watched the MDS, whereas others only glanced at it briefly or did not look at it at all. Viewing duration and whether each respondent actually looked at the MDS were not recorded. After passage, another staff member behind the area encouraged visitors to complete the questionnaire on an iPad, but participation was voluntary. Afterward, the staff provided additional explanations of the experiment's purpose as needed, and the experiment concluded. Visitors were instructed not to complete the questionnaire more than once. No compensation was provided.
The 11 rating-scale items were answered on a five-point Likert scale ranging from 1 (strongly disagree) to 5 (strongly agree), with a midpoint of 3 (neither agree nor disagree).

\section{Structural Associations among Consistency, Presence, and Restoration}\label{sec:validation-model}

This section examines whether the hypothesized MDS restoration model is compatible with participants' responses, using all responses from the eight nature-video configuration--spacing days (\textit{N} = 1025), excluding the advertisement condition.
As described in Sec. \ref{sec:experimental-design}, the dataset includes responses obtained under heterogeneous real-world conditions, in which both video-presentation factors and participants' individual circumstances varied. This analysis therefore aims to characterize the covariance structure of participants' responses pooled across these diverse situations. 

Following the recommended two-step approach \cite{anderson1988structural}, we first evaluated the measurement model using confirmatory factor analysis (CFA) to ensure adequate representation of the latent constructs. We then examined the structural relationships among the latent variables using structural equation modeling (SEM).
In both analyses, the 11 five-point items were treated as ordered categorical indicators, and models were estimated using mean- and variance-adjusted weighted least squares (WLSMV) \cite{asparouhov2010simple}. CFA was conducted in JASP 0.96.0 and SEM in lavaan 0.6-21 in R 4.5.2. 

\subsection{Evaluating the measurement model using confirmatory factor analysis}\label{sec:measurement-model}

\begin{table}[h]
    \centering
    \caption{Results of confirmatory factor analysis (CFA)}
    \begin{tabular}{ccc}
        Factor & Item & Standardized loading \\
        \midrule
        \multirow{2}{*}{Consistency} & continuous & 0.579 \\
         & expanding & 0.851 \\
        \midrule
        \multirow{3}{*}{Presence} & being-there & 0.771 \\
        & easy-to-walk & 0.606 \\
        & existing & 0.787 \\
        \midrule
        \multirow{6}{*}{Restoration} & calm & 0.741 \\
        & fine & 0.834 \\
        & restored & 0.859 \\
        & focused & 0.749 \\
        & stress-free & 0.821 \\
        & clear & 0.797
    \end{tabular}
    \label{tbl:CFA}
\end{table}

CFA was conducted to test the hypothesized three-factor model consisting of consistency, presence, and restoration. All factor loadings were significant ($p<$0.001; Table \ref{tbl:CFA}), supporting the association of the items with their intended constructs. The WLSMV mean- and variance-adjusted test statistic was $\chi^2(41)$ = 369.68, $p<$0.001. The conventional scaled fit indices were CFI = 0.975, TLI = 0.967, and RMSEA = 0.088 (90\% CI [0.080, 0.097]), with SRMR = 0.030, whereas the robust fit indices were CFI = 0.951, TLI = 0.934, and RMSEA = 0.085. The relatively high CFI values and low SRMR support using the three-factor structure as a working measurement model, while the robust TLI below 0.95 and RMSEA values above 0.08 indicate room for refinement. The limitations of using only two consistency items and the need for additional validated items are discussed in Sec. \ref{sec:limitation-measurement}.

\subsection{Structural equation modeling (SEM)}\label{sec:sem}

Using a three-factor measurement model comprising consistency, presence, and restoration, examined in Sec. \ref{sec:measurement-model}, we conducted SEM to examine the structural relationships among these latent factors. The hypothesized model (Model (1) in Table \ref{tbl:plausible-model}) was evaluated and compared with four alternative models.

\begin{figure}[!htbp]
    \centering
    \includegraphics[width=0.6\linewidth]{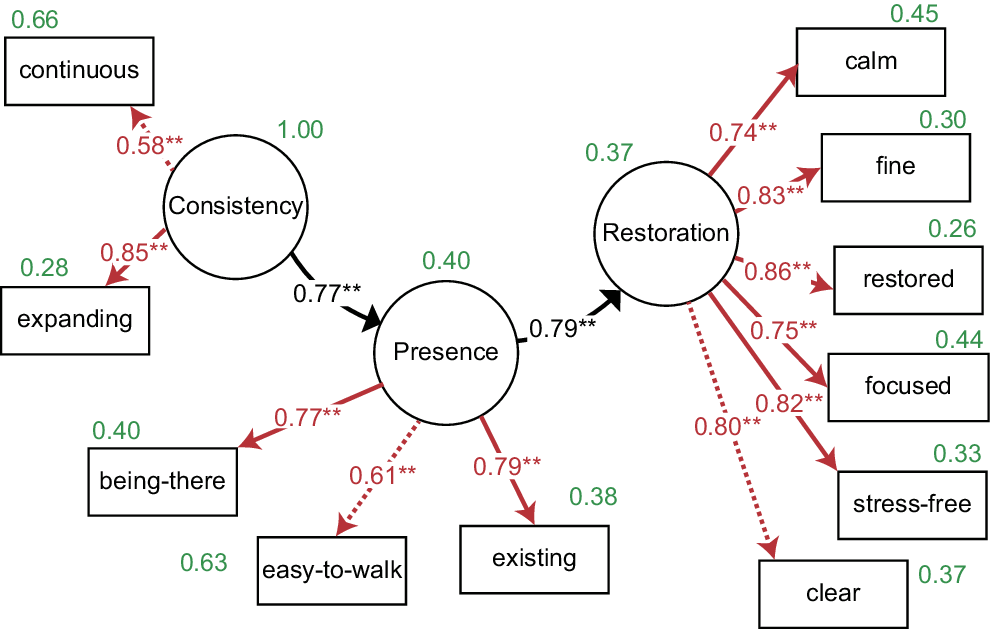}
    \caption{The path diagram estimated by SEM for the hypothesized model (C$\rightarrow$ P $\rightarrow$ R; Model (1) in Table \ref{tbl:plausible-model}). The green, black, and red values indicate the variance of the variable itself, the standardized path coefficient, and the standardized factor loading, respectively. Dashed red arrows indicate marker indicators. ** indicates significance at $p < 0.001$. Single-headed arrows denote the theoretically specified paths; the cross-sectional analysis estimates associations (Sec. \ref{sec:alt-models})}
    \label{fig:3factor-model}
\end{figure}

The standardized path coefficients were large and significant (C$\rightarrow$P: $\beta$ = 0.775; P$\rightarrow$R: $\beta$ = 0.792; both $p<$0.001; Fig. \ref{fig:3factor-model}), indicating strong associations consistent with the hypothesized structure. The model showed reasonable but imperfect fit: WLSMV $\chi^2(42)$ = 360.41, $p<$0.001; conventional scaled CFI = 0.976, TLI = 0.969, and RMSEA = 0.086 (90\% CI [0.078, 0.094]); SRMR = 0.030; and robust CFI = 0.951, TLI = 0.936, and RMSEA = 0.083. The lower adjusted $\chi^2$ relative to the CFA reflects differences in the mean- and variance-adjustment rather than improved fit (Online Resource 1, Section S3). The following comparison examines which aspects of the proposed structure can be distinguished from alternative models.

In summary, the estimated associations were strong for both consistency–presence and presence–restoration, indicating that participants who perceived greater consistency across displays also tended to report stronger presence, which in turn was associated with higher restoration ratings. The following comparison examines which aspects of the proposed structure can be distinguished from alternative models.

\subsection{Comparison with alternative SEM models}\label{sec:alt-models}

We compared the hypothesized model with four theoretically plausible alternatives (Table \ref{tbl:plausible-model}) to examine whether adding a direct consistency--restoration path improved fit and whether the data distinguished among different structural specifications. All five models used the same three-factor measurement model.

\begin{table}[tbp]
    \centering
    \caption{SEM results for the hypothesized model (1) and four alternative models (2--5). C, P, and R denote consistency, presence, and restoration, respectively. The symbols $\rightarrow$ and $\leftrightarrow$ denote regression paths and covariance relationships, respectively. The three columns on the right indicate the standardized coefficients of each path.}
    \footnotesize
    \setlength{\tabcolsep}{2.4pt}
    \begin{tabular}{c>{\centering\arraybackslash}p{2.0cm}c>{\centering\arraybackslash}p{1.3cm}ccccccccc}
         & & & & \multicolumn{3}{c}{Scaled$^{b}$} & \multicolumn{3}{c}{Robust$^{c}$} & \multicolumn{3}{c}{Path$^{d}$} \\ \cmidrule(lr){5-7}\cmidrule(lr){8-10}\cmidrule(lr){11-13}
         & Model & $df$ & WLSMV $\chi^2(df)^{a}$ & CFI & TLI & RMSEA & CFI & TLI & RMSEA & C-P & P-R & C-R \\ \cmidrule(lr){1-10}\cmidrule(lr){11-13}
         (1)&C$\rightarrow$P$\rightarrow$R & 42 & 360.41 & 0.976 & 0.969 & 0.086 & 0.951 & 0.936 & 0.083 & 0.775 & 0.792 & - \\

        (2)&C$\leftrightarrow$P$\rightarrow$R & 42 & 360.41 & 0.976 & 0.969 & 0.086 & 0.951 & 0.936 & 0.083 & 0.775 & 0.792 & - \\ \cmidrule(lr){1-10}\cmidrule(lr){11-13}

        (3)&C$\rightarrow$P$\rightarrow$R, C$\rightarrow$R & 41 & 369.68 & 0.975 & 0.967 & 0.088 & 0.951 & 0.934 & 0.085 & 0.775 & 0.794 & -0.003 \\

        (4)&C$\leftrightarrow$P, P$\rightarrow$R, C$\rightarrow$R & 41 & 369.68 & 0.975 & 0.967 & 0.088 & 0.951 & 0.934 & 0.085 & 0.775 & 0.794 & -0.003 \\

        (5)&C$\rightarrow$P, P$\leftrightarrow$R, C$\rightarrow$R & 41 & 369.68 & 0.975 & 0.967 & 0.088 & 0.951 & 0.934 & 0.085 & 0.775 & 0.635 & 0.613
    \end{tabular}

    \vspace{4pt}
    \begin{minipage}{\linewidth}
    \footnotesize
    \raggedright
    $^{a}$Mean- and variance-adjusted WLSMV $\chi^2$ ($N$ = 1025; all $p<$0.001). Nested comparisons use the Satorra (2000) scaled difference test \cite{satorra2000scaled}, not direct subtraction of these model-specific statistics (Sec. \ref{sec:alt-models}; Online Resource 1, Section S3).\\
    $^{b}$Conventional scaled fit indices. $^{c}$Robust categorical-data fit indices \cite{savalei2021improving}; SRMR = 0.030 for all models.\\
    $^{d}$Standardized regression coefficients for $\rightarrow$; latent C--P correlations in Models (2) and (4) and a residual P--R correlation in Model (5) for $\leftrightarrow$. Correlations and regression coefficients are not directly comparable.
    \end{minipage}
    \label{tbl:plausible-model}
\end{table}

The five models form two groups with identical fit within each group. Models (1) and (2) differ in whether consistency predicts presence or covaries with it, but they are statistically equivalent \cite{maccallum1993problem}. Models (3)--(5) are also statistically equivalent to one another and yield the same fit as the three-factor CFA. Details of the model constraints and nested comparisons are provided in Online Resource 1 (Section S3).

A scaled difference test comparing Model (1) with Model (3), which additionally includes a direct path from consistency to restoration, found no evidence that adding this path improved model fit, $\Delta\chi^2(1)$ = 0.002, $p$ = 0.962. The estimated direct path in Model (3) was close to zero ($\beta$ = $-$0.003, $p$ = 0.962), indicating little unique association between consistency and restoration after accounting for presence in this sample. The alternative models are either statistically equivalent to the proposed model or differ by a single constraint that was not rejected, so model fit does not identify path direction or causal mediation. Taken together, these comparisons support using the proposed model as a parsimonious, theoretically motivated description of the observed associations.

\subsection{Summary and considerations}
The analysis identified strong associations among perceived consistency, presence, and restorative responses in a museum field setting with participants spanning a wide age range. Within the model, the association between consistency and restoration was largely accounted for by presence. These findings support the three-factor model as a concise framework for describing MDS-mediated restorative experiences and developing directional hypotheses. The cross-sectional design and statistically equivalent alternatives leave causal direction and mediation unresolved (Sec. \ref{sec:alt-models}).

Experimental or longitudinal studies can build on these associations to test the proposed sequence, alongside further refinement of the measurement model. Measurement-related considerations, including shared method variance and fixed item order, are discussed in Sec. \ref{sec:limitation-measurement}.

\section{Exploratory Comparison of Presentation Configurations}\label{sec:stat-analysis}

The previous section showed that participants' responses were compatible with the hypothesized structure. This section explores whether ratings of the three measures differed across complete presentation configurations, which varied in video ordering, viewpoint motion, and display spacing.
Figure \ref{fig:stats} summarizes the mean and standard deviation of consistency, presence, and restoration for each video configuration. The nature-video data are pooled across the two spacing conditions, whereas AD was presented only under large spacing.
JASP 0.96.0, which internally uses R 4.5.2, was used consistently for the following statistical analyses.

\subsection{Comparison of video and display spacing conditions}\label{sec:stat-analysis-nature}
To compare responses across video configurations and spacing conditions, a two-way ANOVA was conducted with video condition (MP, MR, MS, and FP) and spacing condition as factors, using only the nature-video data ($N=1025$). For consistency, Levene's test was significant ($p<0.05$), suggesting violations of the homogeneity-of-variance assumption. However, because the ratio of the largest to smallest cell-wise variance was modest (1.780; see Table S3 in Online Resource 1), the factorial ANOVA was considered reasonably robust for the main analysis. Additional sensitivity analyses using HC3-robust ANOVA based on the heteroskedasticity-consistent type 3 covariance estimator \cite{MacKinnon1985HC} yielded results largely consistent with those of the factorial ANOVA (Online Resource 1, Section S4, Table S2). Accordingly, this section reports the results of the factorial ANOVA.

\begin{figure}
    \centering
    \includegraphics[width=0.8\linewidth]{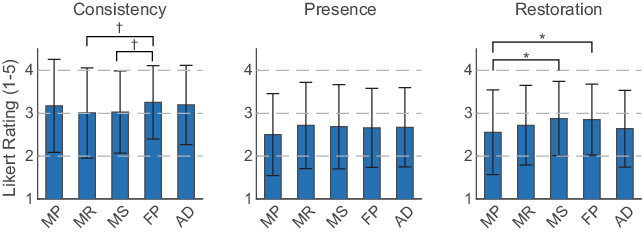}
    \caption{The mean and standard deviation (SD) calculated for each video condition and each factor. For the nature-video conditions, values are pooled across the two spacing conditions; the AD condition includes only the large-spacing condition. * and $\dagger$ indicate significant pairwise differences at $p < 0.01$ and $p < 0.05$, respectively. These pairwise comparisons were conducted only within the nature-video conditions (see Sec. \ref{sec:stat-analysis-nature}). Each nature-video configuration was presented on one weekend day (large spacing) and one weekday (small spacing), so differences may partly reflect day-level factors}
    \label{fig:stats}
\end{figure}

For consistency, a significant main effect of video condition was found ($F$(3,1017) = 4.293, $p = 0.005$, $\eta^2$ = 0.012), whereas neither the main effect of spacing condition ($F$(1,1017) = 0.349, $p = 0.555$, $\eta^2$ $<$ 0.001) nor the interaction was significant ($F$(3,1017) = 1.277, $p = 0.281$, $\eta^2$ = 0.004).

For presence, neither the main effect of video condition ($F$(3,1017) = 1.241, $p$ = 0.293, $\eta^2$ = 0.004), the main effect of spacing condition ($F$(1,1017) = 1.207, $p$ = 0.272, $\eta^2$ = 0.001), nor the interaction was significant ($F$(3,1017) = 2.373, $p$ = 0.069, $\eta^2$ = 0.007).

For restoration, a significant main effect of video condition was found ($F$(3,1017) = 4.929, $p$ = 0.002, $\eta^2$ = 0.014), whereas neither the main effect of spacing condition ($F$(1,1017) = 0.098, $p$ = 0.754, $\eta^2$ $<$ 0.001) nor the interaction was significant ($F$(3,1017) = 0.794, $p$ = 0.497, $\eta^2$ = 0.002).

Taken together, the analyses identified modest differences in consistency and restoration across video configurations ($\eta^2\approx0.01$), with no significant differences in presence or between spacing conditions. The configuration differences provide initial indications of variation in viewers' experiences. Because spacing coincided with weekdays versus weekends, its independent contribution remains unresolved.

Post hoc pairwise comparisons among the video conditions for consistency and restoration were conducted with Bonferroni correction, and the results are shown in Fig. \ref{fig:stats}. For consistency, significant differences were found between MR and FP (\textit{t}(1017) = 3.048, $p$ = 0.014) and between MS and FP (\textit{t}(1017) = 2.793, $p = 0.032$), indicating that consistency ratings were higher in FP than in MR and MS. For restoration, significant differences were found between MP and FP (\textit{t}(1017) = 3.267, $p=0.007$) and between MP and MS (\textit{t}(1017) = 3.339, $p=0.005$). No significant differences were observed for pairs not reported here.

Restoration ratings were higher in FP and MS than in MP, despite MP being designed to combine viewpoint-motion cues with spatial continuity across displays. This unexpected pattern motivates a closer examination of how viewers experienced the configurations, drawing on the open-ended responses in Sec. \ref{sec:consideration-compare-open-ended}.

\subsection{Comparison with advertisement (AD)}\label{sec:compare-AD}

To compare restorative responses to MDS-mediated nature with those to the conventional advertisement (AD) reference condition,
a one-way ANOVA across the five video conditions (MP, MR, MS, FP, and AD), restricted to the large-spacing condition data ($N=769$), was conducted. No heteroscedasticity correction was applied because Levene's homogeneity test was not significant (\textit{F}(4,764) = 1.96, $p = 0.099$ for consistency, \textit{F}(4,764) = 1.508, $p = 0.198$ for presence, and \textit{F}(4,764) = 1.982, $p$ = 0.095 for restoration).

It revealed significant differences in presence (\textit{F}(4,764) = 2.784, $p$ = 0.026, $\eta^2$ = 0.014) and restoration (\textit{F}(4,764) = 2.647, $p$ = 0.032, $\eta^2$ = 0.014), whereas no significant difference was found for consistency (\textit{F}(4,764) = 1.066, $p$ = 0.372, $\eta^2$ = 0.006). As post hoc tests, Dunnett's tests were conducted for presence and restoration using the AD condition as the control. However, no significant differences were found between the AD condition and any other condition.

The nature-video configurations did not show significantly higher restoration ratings than the AD reference in this study. Generalization is limited because AD was a specific advertisement for the museum, presented only on one weekend day under large spacing, and was not matched to the nature videos in duration, motion, color, or luminance (Sec. \ref{sec:limitation-allocation}). Within this exploratory study, the comparison nevertheless provides an informative reference for how the tested nature videos were rated relative to existing signage content at the site. The open-ended responses offer further clues for interpreting this result (Secs. \ref{sec:open-text} and \ref{sec:cross-consideration-AD}).

\subsection{Summary and considerations}

The exploratory comparisons revealed modest variation in consistency and restoration across presentation configurations. In particular, FP and MS received higher restoration ratings than MP, although MP was designed to combine viewpoint motion and spatial continuity. Mean restoration ratings remained around or below the neutral midpoint, and the nature-video configurations did not show a significant advantage over the AD reference.

These patterns identify questions for MDS content design: how viewpoint motion and spatial arrangement relate to restorative impressions, and which qualities of the imagery encourage or detract from those impressions. The open-ended responses complement the comparisons by describing experiences that may help explain the moderate ratings and guide future design studies (Sec. \ref{sec:consideration-compare-open-ended}).


\section{Analysis of open-ended responses}\label{sec:open-text}

In this section, we analyzed the open-ended responses using an established qualitative approach based on coding and categorization \cite{Hsieh2005ContentAnalysis,Elo2008ContentAnalysis}.
This analysis aimed to capture and categorize participants' feelings and experiences regarding the nature videos presented across the MDS, including psychological states that may not have been fully captured by the five-point scales, and to explore possible improvements for enhancing restorative experiences with nature videos on MDS.
The 352 open-ended responses collected over the eight days when nature videos were presented on the MDS were used for the qualitative analysis, accounting for 34.34\% of all responses under the nature-video conditions (see Fig. \ref{fig:answer-num}(a)). 

\subsection{Iterative coding and categorization procedure}
Before the main analysis, we shuffled the order of all responses and removed both the video-condition labels and the associated quantitative scores, so that only the open-ended responses were retained for coding, in order to reduce potential bias.
We familiarized ourselves with the data by reading the full corpus repeatedly. Each response was assigned one to three codes, reflecting the fact that individual responses often addressed multiple aspects of the experience rather than a single topic. We then developed higher-level categories that grouped related codes. During both the initial coding and subsequent categorization, we iteratively moved back and forth between the original responses and the constructed coding scheme to ensure that the codes and categories were constructed carefully and comprehensively. We further checked the resulting categories by examining their correspondence with the original responses; when participant responses did not fit existing categories, we revisited and refined the category structure. One author conducted the initial coding and produced a preliminary category list. The mapping between the categories and the original data was then reviewed by at least one other author, and any ambiguous cases were discussed further to refine the coding decisions and define the final set of categories.

\subsection{Results of the open-ended response analysis}\label{sec:open-text-result}

\begin{table}[]
    \centering
    \caption{Categories and subcategories for open-ended responses. Commas separate subcategories; slashes are part of subcategory names and do not indicate separate subcategories.}
    \begin{tabular}{ll}
         Category & Subcategory \\
         \midrule\midrule
         Restoration & Fascination, Healing, Being away, Nostalgia \\
         \midrule
         \multirow{2}{*}{Spatiality} & Immersion, Visual impact,  \\
         & Walking/movement compatibility, Continuity \\
         \midrule
         \multirow{2}{*}{Discomfort} & Unnaturalness, Unpleasantness, Motion sickness, \\
         & Low resolution, No effect \\
         \midrule
         \multirow{2}{*}{Non-spatiality} & Discontinuity, Walking/movement incompatibility, \\
         & Inattention \\
         \midrule
         Constructive & Larger screen / longer time, \\
         feedback & Improvement suggestion, Association/application \\
         \midrule
         Unclear & Difficult/unclear, Others \\
    \end{tabular}
    \label{tbl:categories}
\end{table}
\begin{figure}
    \centering
    \includegraphics[width=0.7\linewidth]{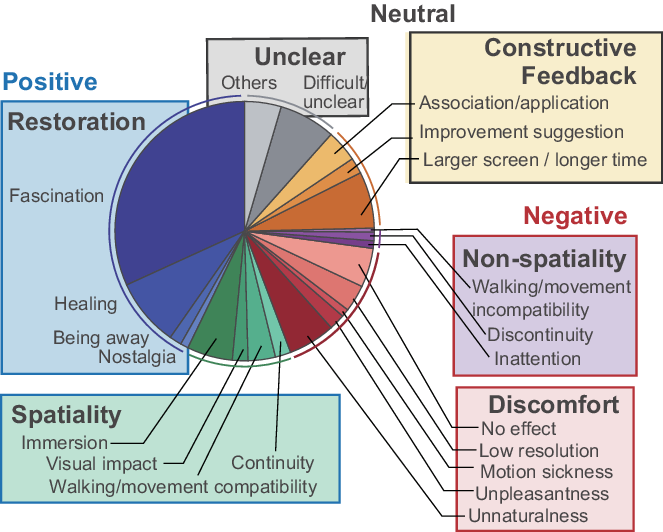}
    \caption{Proportions of categories and subcategories obtained through iterative coding and categorization of the open-ended responses across the eight nature-video conditions}
    \label{fig:desc-ratio}
\end{figure}

The categories and subcategories used to classify the open-ended responses are shown in Table \ref{tbl:categories}, and their proportions are shown in Fig. \ref{fig:desc-ratio}.
The classification criteria for each subcategory are shown in Table S4; condition-wise weighted proportions are provided in Table S5 and Fig. S4 (Online Resource 1, Section S7).
Of all responses, 76.99\% were assigned one subcategory, 20.17\% were assigned two subcategories, and 2.84\% were assigned three subcategories.
When multiple subcategories were assigned to a single comment, the score for that comment was divided equally among the assigned subcategories. For example, if three subcategories were assigned to one comment, each subcategory received a score of 1/3. For each subcategory, the sum of the scores was calculated and then divided by the total number of comments to obtain the percentage of that subcategory relative to all comments. All category and subcategory percentages below refer to shares of this weighted coding score. 

The categories were broadly grouped into three types: positive, negative, and neutral responses, as shown in Fig. \ref{fig:desc-ratio}. These groups are discussed in the following subsections.

\subsubsection{Positive responses}
\qualparagraph{Restoration} This category accounted for 42.80\% of the total weighted coding score across the 352 open-ended comments. Fascination was the largest subcategory (31.82\%), encompassing positive reactions such as beauty, excitement, interest, and attractiveness (Table S4 in Online Resource 1). Participants described the videos as {\itshape \enquote{interesting}} or {\itshape \enquote{wonderfully mysterious,}} and some expressed a desire to watch them again. Healing (8.48\%) captured explicit accounts of calmness and refreshment, while Being away (1.52\%) and Nostalgia (0.99\%) reflected other restorative or personally meaningful impressions.

Among participants who left comments, MDS-mediated nature was described as attractive and, in some accounts, calming or refreshing. Fascination and Being away provide thematic links to Attention Restoration Theory (ART) \cite{kaplan1989experience,kaplan1995restorative}. Here, the Fascination code captures general interest and attraction and may not correspond exactly to ART's soft fascination, which involves effortless attention that leaves room for reflection.

\qualparagraph{Spatiality}
This category accounted for 12.97\% of the weighted coding score. Immersion was the most frequent subcategory (5.68\%), encompassing comments about spaciousness and a sense of being present in the depicted scene. Walking/movement compatibility accounted for 3.41\% and included comments describing a sensation of flying or a stronger sense of being in the scene when walking speed matched the camera movement. Visual impact and Continuity each accounted for 1.94\%.

These accounts describe experiences of entering or moving through the depicted natural environment, complementing the questionnaire measures of presence with viewers' descriptions of spatial immersion and movement compatibility.

\subsubsection{Negative responses}\label{sec:open-text-negative}

\qualparagraph{Discomfort}
Negative impressions related to discomfort accounted for 17.09\% of the weighted coding score. These comments primarily concerned unnatural colors or an artificial appearance (Unnaturalness: 5.87\%), a lack of noticeable psychological effects due partly to limited exposure or attention (No effect: 4.88\%), and poor or blurry image quality (Low resolution: 3.36\%). Less frequent concerns included unpleasant impressions caused by high-contrast colors or the videos and MDS themselves (Unpleasantness: 1.99\%) and visually induced motion sickness (Motion sickness: 0.99\%).

Participants linked discomfort to low video resolution, unnatural colors, and visually induced motion sickness. Their accounts identify specific aspects of the viewing experience to address in future content design and evaluation.

\qualparagraph{Non-spatiality}
This category accounted for only 2.51\% of the weighted coding score and comprised Discontinuity (1.09\%), Inattention (0.95\%), and Walking/movement-incompatibility (0.47\%). The latter, referring to video motion perceived as faster than the viewer's walking speed, appeared only alongside other subcategories, suggesting that it was not a salient standalone impression.

Although uncommon, these comments highlight discontinuity, inattention, and movement mismatch as aspects of the experience that warrant consideration when imagery is distributed across separated displays within a limited part of the viewer's visual field.

\subsubsection{Neutral responses}
\qualparagraph{Constructive feedback}
This category accounted for 13.02\% of the weighted coding score. The largest subcategory was Larger screen / longer time (7.05\%), which included requests for a more immersive presentation using larger, additional, or surrounding displays, as well as longer viewing durations. Association/application comments (3.98\%) suggested potential uses of the MDS in settings such as stations or as ambient displays, while Improvement suggestion comments (1.99\%) proposed changes to the display arrangement or the addition of sound.

Although these comments cannot be clearly interpreted as either positive or negative, they can be understood as reflecting interest in the videos or the technology, personal associations with the viewers' own experiences, and further expectations or desires arising from the experience.

\qualparagraph{Unclear}
This category accounted for 11.60\% of the weighted coding score. The largest subcategory was Difficult/unclear (7.05\%), which included difficulty understanding the intention, content, or spatial arrangement of the videos. The remaining comments were categorized as Others (4.55\%) and consisted mainly of responses indicating that the participant had nothing particular to report or containing no interpretable content.

Responses categorized as Difficult/unclear may indicate that the brief and visually limited MDS experience was difficult for some viewers to interpret.

\subsection{Summary and considerations}\label{sec:open-text-consideration}
The open-ended responses revealed a range of experiences that complemented the rating-scale results. Positive comments encompassed fascination, restorative feelings, and spatial immersion, while negative and ambivalent accounts highlighted perceived unnaturalness, image quality, and motion-related discomfort. Commenting was optional, and the category proportions summarize the content of responses from participants who chose to comment, using shares of the weighted coding score. They do not indicate how common each experience was among all participants (Sec. \ref{sec:open-text-bias}). Their implications for refining the nature imagery, restoration model, and questionnaire are discussed in Sec. \ref{sec:consideration}.

The following discussion develops three aspects of these findings: their thematic connections with ART, the coexistence of positive and negative impressions, and the scope of experiences captured through optional comments.

\subsubsection{ART-related responses}\label{sec:discussion-ART}

Several subcategories may be interpreted as thematically related to key components of ART \cite{kaplan1989experience,kaplan1995restorative}. Walking/movement compatibility (3.41\%) may reflect a sensorimotor form of {\itshape Compatibility}, whereas Larger screen / longer time (7.05\%) may indicate a desire for a greater sense of {\itshape Extent}. Together with Fascination and Being away, these thematic links identify aspects of MDS-mediated experience that could be examined with ART-informed measures in future studies.

\subsubsection{Ambivalent responses}
Interestingly, some open-ended responses expressed both positive (+) and negative (-) feelings. For example, the following comment was categorized as Unnaturalness (-), Healing (+), and Unpleasantness (-):
\begin{displayquote}
{\itshape \enquote{It may have been live-action footage, but it felt like an artificial video. It reminded me of Lassen's dolphin paintings. I did not dislike it, but while it made me feel relaxed, I also felt a slight sense of tension.}}
\end{displayquote}
This suggests that, although the AI-generated nature video partially elicited relaxation for this participant, its perceived unnaturalness may have hindered its acceptance as nature.

Another comment was categorized as Healing (+), Nostalgia (+), and Unpleasantness (-):
\begin{displayquote}
{\itshape \enquote{I felt relieved because it reminded me of the sea in my hometown, but I also felt a slight stress when I returned to the reality of being surrounded by inorganic objects in the city. This is a valuable study. I support it.}}
\end{displayquote}
This comment reflects an ambivalent experience: the imagery evoked nostalgia and relief, while returning attention to the surrounding urban environment was described as stressful. This may have been partly due to the experience occupying only a limited part of the visual field and lasting only a short time.

Together, these comments point to several possible design challenges. To enable the general public to more readily experience restorative benefits from nature videos, it may first be necessary to improve their visual quality by addressing issues such as perceived unnaturalness and insufficient resolution. The nature imagery presented on the MDS should also be designed to blend more naturally into the urban landscape, thereby reducing the perceptual gap between the real environment and the depicted natural scenes. However, if the imagery blends too seamlessly into its surroundings, it may recede into the background and fail to attract attention. The design must therefore balance integration into the urban environment with sufficient visual salience to maintain its potential restorative impact.

\subsubsection{Self-selection in open-ended responses}\label{sec:open-text-bias}
In the descriptive comparison, participants who provided open-ended responses had higher mean questionnaire scores in some conditions, but this pattern was not consistent across conditions and constructs (Online Resource 1, Section S6, Fig. S3). The comments therefore offer insight into the range of experiences among those who chose to elaborate on their impressions, with self-selection limiting generalization to the full sample.

\section{Cross-cutting Considerations}\label{sec:consideration}

This study involved three data analyses: CFA and SEM for examining the hypothesized MDS restoration model (Sec. \ref{sec:validation-model}), exploratory statistical comparisons across presentation configurations and spacing conditions (Sec. \ref{sec:stat-analysis}), and qualitative analysis of participants' open-ended responses (Sec. \ref{sec:open-text}). This section discusses several cross-cutting considerations derived from the multiple analyses reported above.

\subsection{Cross-considerations from the SEM and the comparative experimental results}\label{sec:consideration-SEM-compare}

The SEM identified strong associations linking consistency with presence and presence with restoration (Sec. \ref{sec:validation-model}). This framework helps organize the condition-wise patterns in Fig. \ref{fig:stats} and identify questions raised by differences across configurations. The SEM describes associations in pooled individual responses, whereas the comparisons concern condition means obtained on different days; this discussion therefore connects the two levels descriptively.

MP tended to show relatively high consistency but low presence, and its restoration score was significantly lower than those of MS and FP. MR, in contrast, showed significantly lower consistency than FP, while its presence score was similar to that of FP, but its restoration score also tended to be lower than that of FP. MS showed a pattern similar to MR in terms of consistency and presence, with significantly lower consistency than FP but a comparable presence score. However, unlike MR, MS showed a restoration score comparable to that of FP, and both MS and FP showed significantly higher restoration scores than MP.

The condition-wise patterns reveal variation beyond the simple ordering of the three constructs: for example, MS showed lower consistency than FP but comparable restoration. This contrast motivates examining additional dimensions of the viewing experience alongside consistency and presence. The open-ended responses suggest several such dimensions (Sec. \ref{sec:possible-improvement}), while day-level differences remain an alternative explanation for the configuration-wise patterns.

\subsection{Cross-considerations from the comparative experimental results and open-ended responses}\label{sec:consideration-compare-open-ended}

\subsubsection{Consistency of positive categories with the MDS restoration model}

Several categories identified in the open-ended responses overlapped with the items included in the proposed MDS restoration model. Continuity reflected perceived consistency across the displays, while Immersion and Walking/movement compatibility captured aspects of presence. Comments categorized as Healing also directly reflected perceived restorative feelings. Other positive subcategories, including Fascination, Being away, Nostalgia, and Visual impact, may be related to restorative or spatial experiences, although they do not map uniquely onto the three constructs. These findings suggest that the three constructs in our model captured relevant aspects of the experiences described by commenting participants. At the same time, the open-ended responses raised additional factors not explicitly included in the questionnaire, such as motion sickness, perceived unnaturalness, and limited visual quality, which may inform future extensions of the model.

\subsubsection{Viewpoint motion and motion-related discomfort}\label{sec:motion-sickness-nature}
Although MP was expected to promote presence and restoration through spatial coherence, its presence ratings did not differ significantly from those of the other conditions, and its restoration ratings were significantly lower than those of MS and FP (Sec. \ref{sec:stat-analysis}). A small number of open-ended responses mentioned motion sickness in all three moving-viewpoint conditions (weighted proportions: MP, 1.04\%; MR, 2.13\%; MS, 1.01\%), whereas none did in FP (Sec. \ref{sec:open-text}). These comments motivate examining motion-related discomfort, although they were not specific to MP.

The estimated mean optical-flow velocity in the moving-viewpoint videos was 3.42$\pm$1.28$^\circ$/s at a viewing distance of 2.0 m (Sec. \ref{sec:stimuli} and Online Resource 1, Section S1, Fig. S1). Forward viewpoint motion may also make visual motion perceptually salient: radial/expanding flow, which signals self-motion in depth, has been shown to induce stronger vection, motion sickness, and presence than lateral translation \cite{hughes2024factors}.

However, the open-ended responses alone did not fully explain why restoration ratings were lower in MP than in MS. Park et al. \cite{park2022mixing} showed that adding optical flow opposite to virtual visual motion reduced simulator sickness in VR. Drawing on this finding, a post hoc hypothesis is that differently directed optical flows across neighboring displays in MS might partially offset one another and reduce motion-related discomfort \cite{kim2023dynamically,xiao2016augmenting}. 

Building on evidence that calibrated optical flow can support presence \cite{luu2021effects,hughes2024factors}, future studies should manipulate optical-flow direction and test slower velocities while directly measuring discomfort and presence.

\subsubsection{Interpretation of moderate restoration scores}\label{sec:cross-consideration-AD}
Mean restoration ratings were around or below the neutral midpoint of 3 (\enquote{neither agree nor disagree}; Fig. \ref{fig:stats}), and the nature videos used in this study did not yield significantly higher ratings than the AD reference (Sec. \ref{sec:compare-AD}). The open-ended responses described fascination and restorative feelings alongside discomfort (17.09\% of the weighted coding score), including concerns about artificial appearance and low image resolution; some comments combined positive and negative impressions (Secs. \ref{sec:open-text} and \ref{sec:open-text-consideration}). These mixed impressions help interpret the moderate mean ratings. The negative accounts also highlight naturalness, visual quality, and optical-flow design as priorities for further evaluation.

\subsection{Possible improvements to nature-video design, the MDS restoration model, and questionnaire design}\label{sec:possible-improvement}

Integrating the three analyses provides a basis for refining nature-video presentation, the MDS restoration model, and its measurement. The SEM identifies relationships among consistency, presence, and restoration, while the configuration comparisons and open-ended responses draw attention to additional aspects of the experience, including video-content acceptability, motion-related discomfort, and visual quality. We develop these observations into design directions and an extended conceptual model below.

\subsubsection{Nature imagery}\label{sec:nature-imagery} The design of nature imagery could be further improved by matching camera viewpoint motion to viewers' walking speed, increasing image resolution, and expanding the overall visual area by increasing the number of signage displays. In addition, rather than relying solely on AI-generated videos, future studies may need to incorporate footage captured directly in natural environments. In particular, to provide nature imagery with minimal discomfort while enabling precise control of viewpoint motion, it may be beneficial to use wide-angle footage captured by professional camera operators using camera stabilizers.

\subsubsection{Restoration model}\label{sec:adv-restorative-model} We propose a hypothesis-generating extension of the model that incorporates motion sickness, video-content acceptability, and the perceptual factors of naturalness and visual quality (Fig. \ref{fig:teaser}(c)). Derived from the open-ended responses and their interpretation alongside the condition-wise patterns (Sec. \ref{sec:consideration-compare-open-ended}), this extension organizes additional dimensions of MDS-mediated experience for future investigation.
The extension hypothesizes a link between motion sickness and presence to examine whether motion-related discomfort interferes with the sense of being in the depicted environment (Sec. \ref{sec:motion-sickness-nature}). Naturalness and visual quality represent relatively immediate perceptual evaluations of the videos and are proposed to shape video-content acceptability: whether the content is perceived as appropriate, convincing, and acceptable in its setting. In turn, video-content acceptability is hypothesized to relate to presence and restoration, connecting specific perceptual qualities to viewers' broader evaluation of the content and their experience of MDS-mediated nature. The added factors and links remain to be measured and tested.

\subsubsection{Questionnaire design}\label{sec:adv-questionnaire}
The questionnaire design should also be refined in future studies to test this extended model directly. First, items related to motion sickness should be included to explicitly examine whether discomfort caused by moving viewpoints reduces presence and, consequently, restorative experiences. Second, items assessing perceptions of naturalness and visual quality should be developed, together with items measuring video-content acceptability as an evaluative response distinct from presence and restoration.
Furthermore, given that some open-ended responses were thematically consistent with components of Attention Restoration Theory (ART) \cite{kaplan1989experience,kaplan1995restorative} (see Sec. \ref{sec:discussion-ART}), the restoration-related questionnaire items may need to be designed with reference to established ART-based measures \cite{hartig1997measure,pasini2014measure}. These refinements would allow future studies to quantitatively evaluate the role of each factor in the extended MDS model and further examine restorative experiences elicited by nature videos presented on MDS.

To keep response burden manageable when adding these measures in public settings, future studies could randomly assign different subsets of items to participants. This would broaden the measures covered by the study while retaining a brief questionnaire for each viewer.

\section{Limitations}\label{sec:limitation}

This study has several limitations, which are grouped below according to their implications for interpretation.

\subsection{Unbalanced condition allocation and confounding with day and weather}\label{sec:limitation-allocation} 
Display spacing was fully confounded with day type (large on weekends, small on weekdays), and each configuration--spacing combination was tested on a single day. Weather was also unbalanced: MP, MR, and MS under small spacing were tested on rainy days, and all others on sunny days. These comparisons therefore cannot separate presentation effects from day-specific visitor flow, crowding, and composition.
Neither the lack of a clear weather/day-type pattern in the scores (Online Resource 1, Section S5, Fig. S2) nor the nonsignificant spacing result rules out effects of these factors. Future studies should replicate each combination across randomized or counterbalanced days, or alternate conditions within days.

The AD reference condition was presented only under large spacing on a single weekend day. The museum's existing advertisement was not matched to the nature videos in content, motion, color, luminance, or duration. AD therefore provides a reference for these specific presentations under large spacing; this comparison does not establish the relative restorativeness of nature videos and advertisements in general. In future studies, presenting reference content perceptually matched to the nature videos at both spacing levels would enable a more balanced comparison.

\subsection{Prompted viewing and voluntary participation}\label{sec:limitation-participation} The experimental area included other attractions, and most visitors passed through without verbal prompts in the preliminary experiment (Sec. \ref{sec:experimental-setup}). Staff therefore actively encouraged viewing and questionnaire participation. This prompted experience differed from ordinary incidental exposure to public signage and may have increased attention and introduced demand characteristics. The museum setting also differs from intended urban settings such as stations in visitors' motivations, expectations, time pressure, and crowding; generalization therefore requires studies in those settings.

Participation was voluntary, potentially favoring respondents with greater interest in or more positive attitudes toward the MDS. Of nature-video participants, 34.34\% provided comments; their mean scores were higher in some conditions, but not consistently across conditions or constructs (Online Resource 1, Section S6, Fig. S3). Repeat responses could not be identified despite single-response instructions, and returning visitors may have seen other configurations. The broad age range also warrants attention to questionnaire comprehension and age-appropriate consent.



\subsection{Display scale and technical limitations}\label{sec:limitation-technical} The MDS setup consisted of only four 55-inch displays, whereas public urban MDS installations often use more displays with larger screen sizes, such as 60--70-inch displays. Future studies should examine impressions of more prominent urban signage by increasing both the display size and the number of displays.

Although the videos presented on the signage were designed to loop every 24 s, the actual loop duration was approximately 21--23 s, as observed in the supplementary video (Online Resource 2), indicating playback at about 104--114\% of the intended speed (approximately 25--27 fps). This variation was likely caused by the rendering system, which attempted to draw frames as quickly as possible, and possible performance instability of the PC driving the four large signage displays. Because playback speed directly affects optical flow, the actual visual-motion intensity may have been approximately 4--14\% higher than the values estimated in Online Resource 1 (Section S1), which assumed 24 fps. This variation is small relative to the difference between the fixed- and moving-viewpoint videos (0.77 vs. 3.42$^\circ$/s), but frame timing was not logged, so it is unknown whether playback speed differed across days or configurations. Future studies should use more controlled playback systems, such as dedicated signage players, and log presentation timing.

\subsection{Questionnaire design and SEM interpretation}\label{sec:limitation-measurement}

All constructs were assessed at a single time point using the same questionnaire, with items presented in the fixed order of consistency, presence, and restoration. Shared response tendencies and item-order effects may have contributed to the observed associations. To keep the questionnaire brief for museum visitors, we prioritized the six-item assessment of restoration and limited the measurement of consistency to two author-developed items. This may have limited the content coverage and reliability of the consistency factor, as well as the stability of the measurement model. 

The SEM describes associations and does not establish the causal sequence from consistency through presence to restoration, as also illustrated by the alternative-model comparisons in Sec. \ref{sec:alt-models}. Future studies can build on these associations by developing and validating additional consistency items, counterbalancing or randomizing item order, and using experimental manipulations or repeated measurements over time to test the proposed directional relationships.

\section{Conclusion}
We conducted a nine-day exploratory field study in a free-admission area of a public museum, collecting 1,263 questionnaire responses to examine experiences of nature videos presented across distributed displays. Three complementary analyses provided empirical insights into MDS-mediated restorative experiences. First, SEM identified strong associations among perceived consistency, presence, and restorative responses, consistent with the theoretically motivated model. Second, exploratory comparisons revealed modest differences in consistency and restoration ratings across presentation configurations, with FP and MS receiving higher restoration ratings than MP. Mean restoration ratings were around or below the neutral midpoint, and the AI-generated nature videos tested here did not yield significantly higher restoration ratings than the advertisement reference. Third, qualitative analysis captured fascination, immersion, and restorative feelings, together with ambivalent responses and concerns about naturalness, visual quality, and motion-related discomfort. The structural associations and configuration-wise patterns should be interpreted within the study's cross-sectional design and day-level confounding.

Together, these findings document the range of experiences reported in response to MDS-mediated nature and offer directions for refining nature videos and questionnaires for public-display settings. They also inform a tentative conceptual extension of the model incorporating perceptual and evaluative factors. Future studies can build on these findings through replicated, counterbalanced presentations, direct measures of viewing behavior and the additional factors, and pre--post or physiological measures of psychological restoration in urban environments.

\backmatter

\bmhead{Supplementary information}
Online Resource 1 (PDF) provides supplementary methods, the questionnaire, additional statistical analyses, and detailed qualitative results (Sections S1--S7). Online Resource 2 (video) shows the AI-generated nature-video stimuli and their presentation on the MDS (Adobe Firefly, Google Veo 3.1; Sec. \ref{sec:stimuli}).

\bmhead{Acknowledgements}
We would like to express our sincere gratitude to the staff of the National Museum of Emerging Science and Innovation (Miraikan) for their invaluable support in organizing and conducting the experimental demonstration. 

\section*{Statements and Declarations}

\begin{itemize}
\item \textbf{Funding:} This work was supported by the Japan Science and Technology Agency (JST), PRESTO Grant Number JPMJPR25IA.
\item \textbf{Competing interests:} The authors declare that they have no known competing financial interests or personal relationships that could have appeared to influence the work reported in this paper.
\item \textbf{Ethics approval and consent to participate:} The experimental protocol was approved by the research ethics committee of NTT, Inc. (approval number: R07-013) and conducted in accordance with the ethical standards of the Declaration of Helsinki. Consent was obtained as described in the Participants subsection of Experimental design.
\item \textbf{Consent for publication:} Not applicable.
\item \textbf{Data availability:} The anonymized questionnaire data that support the findings of this study are available from the corresponding author upon reasonable request.
\item \textbf{Code availability:} The code used in this study is available from the corresponding author upon reasonable request.
\item \textbf{Author contribution:} YM: Conceptualization, Data curation, Formal analysis, Funding acquisition, Investigation, Supervision, Visualization, Writing -- original draft, Writing -- review \& editing. TY: Investigation, Methodology, Writing -- review \& editing. YK: Investigation, Methodology, Writing -- review \& editing. KM: Investigation, Project administration, Writing -- review \& editing.
\item \textbf{Use of generative AI:} The authors used Adobe Firefly (Google Veo 3.1) to generate the nature-video stimuli from source landscape images, as described in Sec. \ref{sec:stimuli}. During manuscript preparation, the authors used GPT-6 Astra (OpenAI) and Claude (Anthropic) to assist with English-language editing and manuscript revision. The authors reviewed and edited the content as needed and take full responsibility for the content of the published article.
\end{itemize}

\bibliography{citation}

\end{document}